\documentclass[transmag]{IEEEtran}
\usepackage{latexsym}
\usepackage{graphicx}
\usepackage{amsfonts,amssymb,amsmath}
\usepackage{hyperref}
\usepackage{cite}
\usepackage{amsmath}
\usepackage{algorithmic}
\usepackage[ruled,vlined]{algorithm2e}
\usepackage{array}
\usepackage[scr=rsfs]{mathalpha}
\usepackage{color}
\usepackage{gensymb}
\usepackage{booktabs}
\usepackage{enumerate}
\usepackage{subfigure}
\usepackage{caption}

\usepackage{makecell}
\usepackage{verbatim}

\newcolumntype{C}[1]{>{\centering\let\newline\\\arraybackslash\hspace{0pt}}m{#1}}

\usepackage{dblfloatfix}
\usepackage{url}

\def\BibTeX{{\rm B\kern-.05em{\sc i\kern-.025em b}\kern-.08em T\kern-.1667em\lower.7ex\hbox{E}\kern-.125emX}}
\begin{document}

\title{5G Network on Wings: A Deep Reinforcement Learning Approach to the UAV-based Integrated Access and Backhaul}

\author{
    Hongyi Zhang\IEEEauthorrefmark{1}, Zhiqiang Qi\IEEEauthorrefmark{2}, Jingya Li\IEEEauthorrefmark{2}, Anders Aronsson\IEEEauthorrefmark{2},\\ Jan Bosch\IEEEauthorrefmark{1}, Helena Holmström Olsson\IEEEauthorrefmark{3} \\
    \IEEEauthorrefmark{1}\textit{Chalmers University of Technology}, Gothenburg, Sweden.\\
    
    \IEEEauthorrefmark{2}\textit{Ericsson Research}, Ericsson.\\

    \IEEEauthorrefmark{3}\textit{Malmö University}, Malmö, Sweden.

}

\IEEEtitleabstractindextext{
\begin{abstract}
Fast and reliable wireless communication has become a critical demand in human life. In the case of mission-critical (MC) scenarios, for instance, when natural disasters strike, providing ubiquitous connectivity becomes challenging by using traditional wireless networks. In this context, unmanned aerial vehicle (UAV) based aerial networks offer a promising alternative for fast, flexible, and reliable wireless communications. Due to unique characteristics such as mobility, flexible deployment, and rapid reconfiguration, drones can readily change location dynamically to provide on-demand communications to users on the ground in emergency scenarios. As a result, the usage of UAV base stations (UAV-BSs) has been considered an appropriate approach for providing rapid connection in MC scenarios. In this paper, we study how to control multiple UAV-BSs in both static and dynamic environments. We use a system-level simulator to model an MC scenario in which a macro BS of a cellular network is out of service and multiple UAV-BSs are deployed using integrated access and backhaul (IAB) technology to provide coverage for users in the disaster area. With the data collected from the system-level simulation, a deep reinforcement learning algorithm is developed to jointly optimize the three-dimensional placement of these multiple UAV-BSs, which adapt their 3-D locations to the on-ground user movement. The evaluation results show that the proposed algorithm can support the autonomous navigation of the UAV-BSs to meet the MC service requirements in terms of user throughput and drop rate. 
\end{abstract}

}

\maketitle

\section{Introduction}
\IEEEPARstart{T}{raditional} cellular infrastructure provides fast and reliable connectivity in most use cases. However, when a natural disaster happens, such traditional wireless base stations (BSs) can be damaged and therefore they cannot provide mission-critical (MC) services to the users in the disaster area. In this context, further enhancements of the cellular networks are needed to enable temporary connectivity and on-demand coverage for MC users in various challenging scenarios.

Vehicular networking can be enabled by various vehicle types including not only cars but also buses, trucks and UAVs. By equipping with a cellular tower and transceiver on a truck or trailer, cell-on-wheels have fewer cruising duration constraints and can transmit with a higher power to provide a relatively large coverage area \cite{shakhatreh2021cell}. However, cell-on-wheel placement may be less flexible for MC operations in rural areas with complex environments, such as forest firefighting, mountain search and rescue. UAV-BS (cell-on-wings) on the other hand, can be deployed in a more flexible and mobile manner. Specifically, UAVs can be used to carry deployable BSs to provide additional or on-demand coverage to users, thanks to their good mobility and higher chances of light-of-sight (LOS) propagation. However, there are a number of challenges when implementing UAV-BS assisted wireless communication networks in practice \cite{Chalng-7470933}\cite{li2021towards}. The system performance and user experience are significantly impacted by the deployment and configuration of UAV-BSs, including the UAV’s 3-D position, operation time, antenna capabilities, transmit power, etc \cite{9605012}. Using wireless backhaul, UAV-BSs can connect to the on-ground BSs (e.g., cell-on-wheels or macro BSs) and be integrated into the cellular system. Hence, it is necessary to jointly optimize the configuration parameters for the access links (between UAV-BS and on-ground users) and the backhaul links (between UAV-BSs and on-ground BSs), when optimizing UAV-BS based wireless communication systems. The optimization problem becomes even more complicated when considering different system loads and user movement on the ground. In some cases where multiple UAV-BSs are needed to cover a wide area, the complexity of providing reliable and scalable backhaul links will further increase.

Despite the fact that there are numerous applications for UAV-based reinforcement learning algorithms, the fundamental drawback of classic RL is its low performance in a changing environment. If the environment changes (the environmental values observed by the agent change), the agent usually has to retrain the entire algorithm to keep up with the environmental changes \cite{dulac2019challenges}\cite{ding2020challenges}. In our case, user mobility would have a major impact on the system performance in terms of MC user throughput and drop rate. As a result, to ensure good service quality, a triggering mechanism needs to be implemented for algorithm adaptation and analysis. The dynamic environment, in this case, indicates that the states (user throughout and drop rate) that the agent observed will vary substantially due to wireless communication environment changes and user movement.

\begin{figure*}[!htp]
\centering
\subfigure[UAV-BS assisted MC scenario]{
\begin{minipage}[b]{0.5\textwidth}
\vspace{-10pt}
\includegraphics[width=\textwidth]{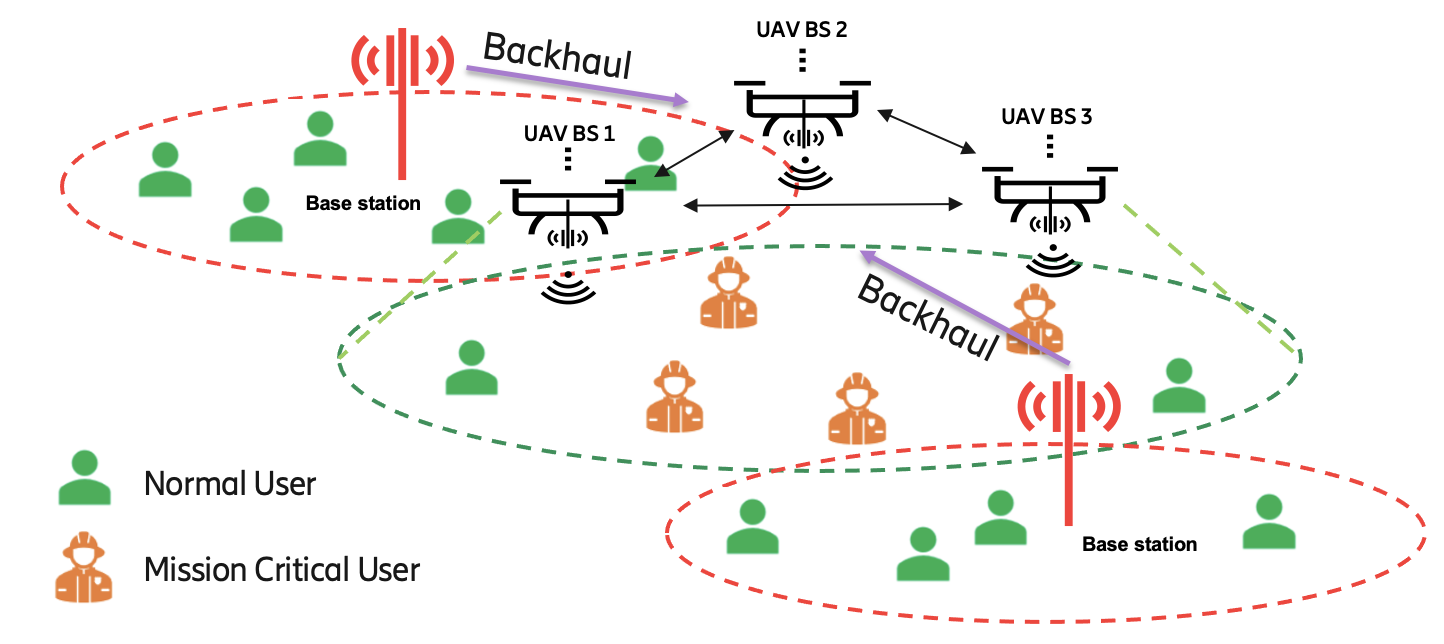} 
\vspace{-10pt}
\label{fig:MCScenario}
\end{minipage}}\subfigure[IAB based TDD duplex pattern]{
\begin{minipage}[b]{0.5\textwidth}
\vspace{-10pt}
\includegraphics[width=\textwidth]{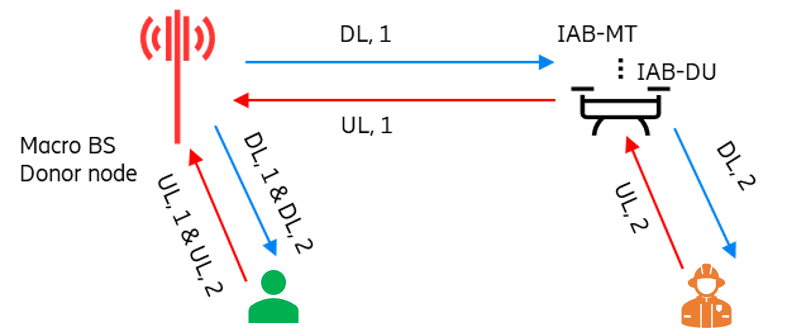}
\end{minipage}}
\subfigure[4 interference cases from the combination of IAB and TDD duplex mode]{
\begin{minipage}[b]{0.8\textwidth}
\vspace{-10pt}
\includegraphics[width=\textwidth]{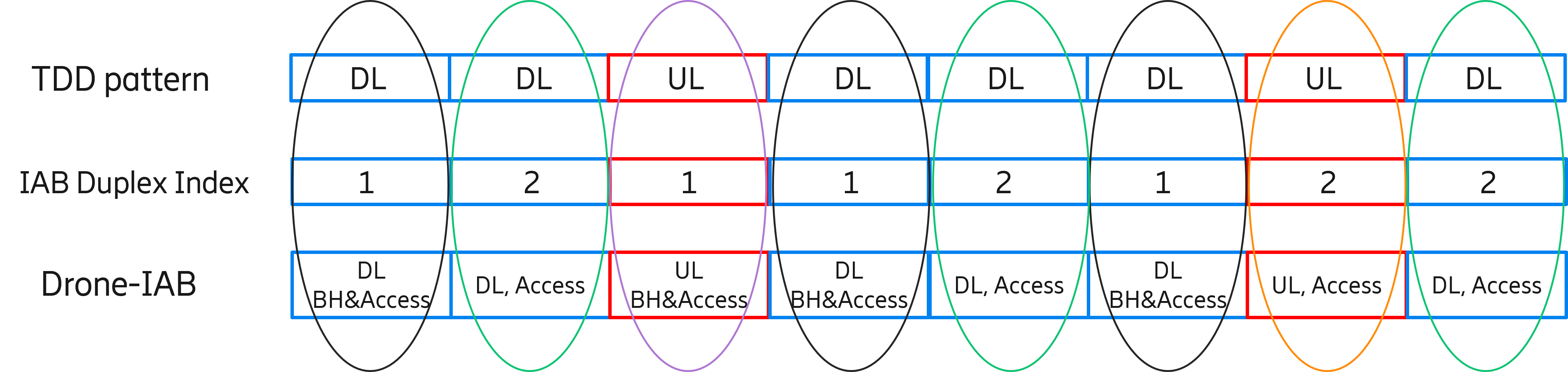}
\vspace{-10pt}
\end{minipage}
}
\vspace{-5pt}
\caption{A UAV-BS assisted wireless network design enabled by a half-duplex IAB operation}
\vspace{-15pt}

\label{fig:Introduction}
\end{figure*}

\subsection{Related Work}
In recent years, UAV-BS assisted wireless communication networks have attracted significant attention from both industry and academia \cite{li20215g, naqvi2018drone, firstnet, merwaday2016improved, ferranti2020skycell}. To guarantee a robust wireless connection between the UAV-BSs and the core network, more and more research work has started working on improving the wireless backhaul link \cite{wang2019deployment,kalantari2017backhaul,cicek2020backhaul,tafintsev2020aerial, Backhaul-9321160}. Authors in \cite{wang2019deployment} assume that all the UAV-BSs are flying at a fixed height, and a robust backbone network among UAV-BSs is guaranteed by ensuring that there is always at least one path between any UAV-BS and a BS on the ground. Then they investigate the rapid UAV deployment problem by minimizing the number of UAVs to provide on-demand coverage for as many users as possible. In \cite{kalantari2017backhaul}, optimal 3-D deployment of a UAV-BS is investigated to maximize the number of connected users with different service requirements by considering the limitation of wireless backhaul links. In \cite{cicek2020backhaul}, the limitation of backhaul and access capacities is also considered, and a heuristic algorithm is proposed to optimize the UAV navigation and bandwidth allocation. Similar to \cite{kalantari2017backhaul}, the authors in \cite{Backhaul-9321160} also investigate a coverage improvement problem enabled by UAV-BS with backhaul limitation but with a machine learning (ML) based solution. 

Enabled by 5G new radio (NR), the integrated access and backhaul (IAB) feature can be applied to wirelessly integrate multiple UAV-BSs to an existing cellular network seamlessly \cite{madapatha2020integrated}. Figure \ref{fig:Introduction} shows an example of UAV-BS assisted network deployment using IAB technology. The macro-BSs who have connections with the core network are serving the normal users, and some of them can also be acting as donor-BSs, who can provide wireless backhaul connections to the flying UAV-BS. Based on the wireless backhaul link, the UAV-BS is acting as an IAB node, which can be deployed at different locations to provide on-demand services to MC users and/or normal users who are out of the coverage of the existing mobile network. To evaluate the performance of the UAV-assisted wireless system enabled by IAB, authors in \cite{tafintsev2020aerial} propose a dedicated dynamic algorithm based on the particle swarm optimization (PSO) method to optimize the throughput and user fairness. By intertwining different spatial configurations of the UAVs with the spatial distribution of ground users, \cite{IAB-8755983} proposes an interference management algorithm to jointly optimize the access and backhaul transmissions. Their results prove that both coverage and capacity can be improved.

Due to the characteristics of revealing implicit features in large amounts of data, the ML methodology draws growing attention and has been extensively applied in various fields. As a sub-field of ML, agent-based reinforcement learning (RL) features in interacting with the external environment and providing an optimized action strategy. Hence, it has been used to solve complicated optimization problems that are difficult to be addressed by traditional methods. As two of the promising technologies for the next-generation wireless communication networks, it is natural to combine ML with deployable UAV-BS to solve high complexity optimization problems \cite{AISurvey-9411810,DL-9353849}.

Specifically, ML is frequently used to solve problems on deployment \cite{ML-Config-8432464,Backhaul-9321160, ML-Deployment-9548327}, scheduling \cite{ML-Scheduling-9127428,ML-Scheduling-9453811,ML-Scheduling-9222519,ML-Scheduling-9526159}, trajectory \cite{ML-Traj-9411725,ML-Traj-8737778,ML-Traj-9475535,ML-Traj-9127423} and navigation \cite{ML-Nav-8600371,ML-Nav-8993742,ML-Nav-8894381,ML-Nav-9354009} in UAV assisted network. In \cite{ML-Config-8432464}, a deep RL-based method is proposed for UAV control to improve coverage, fairness, and energy efficiency in a multi-UAV scenario. To solve the scheduling problem in a high mobility environment, the authors in \cite{ML-Scheduling-9127428} develop a dynamic time-division duplex (TDD) configuration method to perform intelligent scheduling. Based on the experience replay mechanism of deep Q-learning,  the proposed algorithm can adaptively adjust the TDD configuration and improve the throughput and packet loss rate. From the perspective of distributed learning, \cite{ML-Scheduling-9453811} proposes a framework based on asynchronous federated learning in a multi-UAV network, which enables local training without transmitting a significant amount of data to a central server. In this framework, an asynchronous algorithm is introduced to jointly optimize UAV deployment and scheduling with enhanced learning efficiency. 

For ML-based trajectory and navigation, the authors in \cite{ML-Traj-9411725} investigate a trajectory strategy for a UAV-BS by formulating the uplink rate optimization problem as a Markov decision process without user-side information. To enable UAV autonomous navigation in large-scale complex environments, an online deep RL-based method is proposed in \cite{ML-Nav-8600371} by mapping UAV's measurement into control signals. Furthermore, to guarantee that the UAV always navigates towards the optimal direction, authors in \cite{ML-Nav-8993742} enhance the deep RL algorithm by introducing a sparse reward scheme and the proposed method outperforms some existing algorithms.

Additionally, the limited battery life of a UAV restricts its flying time, which in turn affects the service availability that can be provided by the UAV. Therefore, many works have been focusing on designing energy-efficient UAV deployment or configuration schemes either with non-ML \cite{li20215g,EE-7888557,EE-8954798} or ML methodologies \cite{ML-Traj-9127423,EE-ML-9454514,EE-ML-9507262}.  


\vspace{-5pt}
\subsection{Contributions}
\vspace{-5pt}

In this paper, we consider a scenario with multiple macro BSs covering a large area, but due to disaster, one of the macro BSs is damaged, which creates a coverage hole where the first responders execute their MC operations. The deployable UAV-BSs are set up to fill the coverage hole and provide temporary connectivity for these MC users. Compared with the related works and our previous paper\cite{zhang2021autonomous}, we propose in this paper a novel RL algorithm combined with adaptive exploration and value-based action selection algorithms to autonomously and efficiently deploy the UAV-BSs based on the requirements. Furthermore, to extend the algorithm in a scalable manner, a decentralized architecture is proposed for the collaboration of multiple UAV-BSs. More specifically, the contributions of this paper include the following aspects:
\begin{enumerate}[1)]

\item We propose the framework to support applying RL algorithm for the considered use case in an IAB network architecture.

\item We applied two strategies, i.e., adaptive exploration control and value-based action selection for the RL algorithm so that the algorithm itself can adapt to a dynamic environment (e.g., MC user movement) in a fast and efficient way. 

\item We demonstrated deployment in a decentralized method for supporting multiple UAV-BSs deployment to respond to varied industrial scenarios.

\item We validate the proposed RL algorithm in a continuously changing environment with consecutive MC user movement phases. Our results show that the proposed algorithm can create a generalized model and assist in updating the decision-making on UAV-BSs and navigation in a dynamic environment.

\end{enumerate}

The remainder of this paper is structured as follows. Section II introduces the system model considered in this paper. In section III, we propose a framework to enable ML in an IAB network architecture. Section IV discusses our proposed ML algorithm. Section V presents the system-level simulation results and evaluates the proposed RL algorithm. In Section VI, we summarize our findings and discuss future works.

\section{System Model and Problem Formulation}

\begin{figure}[t]
  \begin{center}
    \includegraphics[scale=0.48]{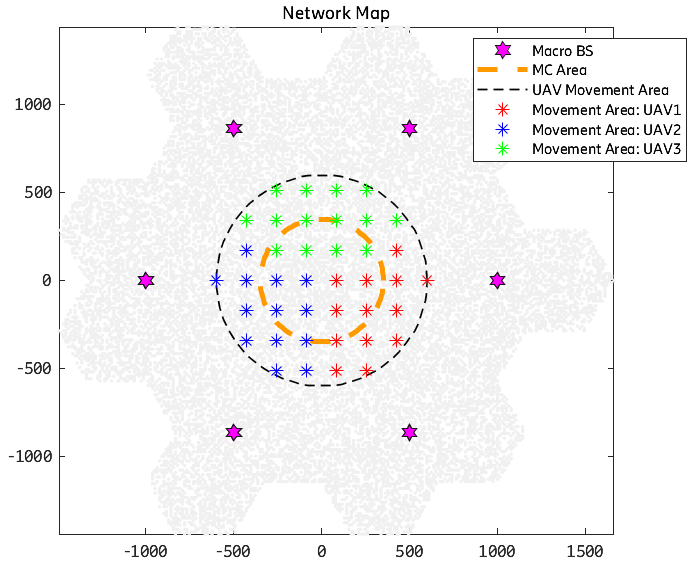}
    \setlength{\abovecaptionskip}{-12pt}
    \caption{System model: UAV-BSs assisted network deployment}
    \vspace{-20pt}
    \label{fig:SimulationScenario}
  \end{center}
\end{figure}

\subsection{System Description}
For the system model, we consider a multi-cell mobile cellular network, consisting of a public network and a deployable network, as shown in Figure \ref{fig:SimulationScenario}. Initially, seven macro-BSs are serving users uniformly distributed in the whole area. However, one of the macro-BSs in the center of the scenario is damaged due to, e.g., a natural disaster that creates a coverage hole. For users in the central emergency area with a predefined radius, they might have very limited or no connectivity with the public network. Hence, multiple UAV-BSs, which are integrated into the public network using IAB technology, can be set up to provide temporary or additional coverage to the users in this emergency area, which is also the research target of this paper. In this paper, the UAV-BSs are limited only to stay at the discrete points indicated by the colored stars in Figure \ref{fig:SimulationScenario}. In the considered scenario, there are two types of users: The users located in the MC area are marked as MC users, while the others are normal users. User equipment (UE), either an MC user or a normal user, can select either a macro-BS or a UAV-BS as its serving-BS, based on the wireless link qualities between the UE and these BSs.



For the traffic pattern design, we apply a dynamic traffic model. All the users are randomly dropped in the scenario. For each time slot, the users are activated with a predefined arrival rate. Only these activated users can be scheduled and initiate fixed-size data transmission based on the link quality (both access and backhaul links) and system load for downlink and uplink, respectively. When the data transmission is completed, the user will leave the system and wait to be activated again. Then the user throughput can be calculated with actually served traffic and consumed time to deliver the traffic.

As mentioned before, the UAV-BSs work as the IAB nodes in the current scenario. They will measure the wireless link to all macro-BSs and select one with the best link quality as their donor-BSs. Once the wireless backhaul link between the UAV-BSs and their donor-BSs are established, the three sectors of the UAV-BSs will share this wireless backhaul link and provide access service to both normal users and MC users. For the users served by the UAV-BSs, the corresponding throughput depends not only on the access link but also on the wireless backhaul link. While selecting the access links, the users with too bad link quality, for instance, below a certain threshold, will be dropped. To reduce the complexity and the load-bearing of the UAV-BSs, it is assumed that the same antenna configuration is applied for both access and backhaul antennas of the UAV-BSs.

The system operates under a TDD model, and the time slot pattern consists of downlink (DL), DL, uplink (UL), and DL, which is repeated with a periodicity of 2 ms \cite{3gpp.38.331}. The system bandwidth is 100 MHz, and it is shared between backhaul and access links. The time slots assigned for UL/DL Access/Backhaul links are shown in Figure \ref{fig:Introduction}(b) and (c). Two full TDD periods are required to cover all eight UL/DL Access/Backhaul combinations, which lead to  four interference cases, denoted as: DL1, DL2, UL1, and UL2. As shown in Figure \ref{fig:Introduction}(b), for the UAV-IAB node, DL1 and UL1 are reserved for backhaul link transmission, while DL2 and UL2 are reserved for providing access services for users. For the donor-BS and all other macro-BSs, all the time slots can be used for access link transmission. 

\subsection{Transmission Model}
For the public network in this paper, we use an urban-macro propagation model \cite{3gpp.38.901}, while a refined aerial model from 3GPP standardization is used for UAV-BS \cite{3gpp.36.777}. It is assumed that the network consists of $D$ macro-BS, $M$ UAV-BSs and $N$ users, denoted by $\mathcal{D}=\{1,2,...,D\}$, $\mathcal{M}=\{1,2,...,M\}$ and $\mathcal{N}=\{1,2,...,N\}$. The whole available bandwidth $W$ is divided into K sub-channels and each one has a bandwidth denoted by $B_k=\frac{W}{K}$. Assuming that the coordinates of the $m^{th}$ UAV-BS and the $n^{th}$ user are $\left( x_m, y_m, h_m \right)$ and $\left( x_n, y_n, h_n \right)$, respectively. Based on 3GPP channel model\cite{3gpp.38.901}, the following formula is applied to represent the probability of LOS propagation between UAV-BS $m$ and user $n$:\\
\begin{equation} \label{eq1}
Pr_{LOS}^{(m,n)} = \begin{cases}
1, &d_{2D}^{(m,n)} \leq d_{2D}^{Th}\\
[ \frac{18}{d_{2D}^{(m,n)} } + \exp (\frac{-d_{2D}^{(m,n)}}{63}) \\\quad \times 
( 1-\frac{18}{d_{2D}^{(m,n)} })] \times \\\quad ( 1 + \frac{5}{4}e^{-6}  \times C^{'}( h_{n} ) {d_{2D}^{(m,n)} }^3 \\\quad \exp{\frac{-{d_{2D}^{(m,n)} }}{150}} ) &d_{2D}^{(m,n)} > d_{2D}^{Th}\end{cases}
\end{equation}
where, \\
\begin{equation} \label{eq2}
C^{'}\left( h_{n} \right) = \begin{cases}
0, &\text{$h_{n} \leq 13$}\\
\left( \frac{h_{n}-13}{10} \right)^{1.5}, &\text{$13 < h_{n} \leq 23$} \end{cases}
\end{equation}
$h_{n}\in \left[h_{n}^{min},h_{n}^{max}\right]$ denotes the height of user n with meter as a unit and $d_{2D}^{(m,n)}= \sqrt{(x_m-x_n)^2+(y_m-y_n)^2}$ is the horizontal distance between UAV-BS $m$ and user $n$. $d_{2D}^{Th}$ is a 2D distance threshold and its value is 18 meters. The path loss between UAV-BS $m$ and user $n$ in the case of LOS propagation and NLOS propagation can also be derived based on \cite{3gpp.38.901}:\\
\begin{equation} \label{eq3}
PL_{LOS}^{(m,n)} = 28 + 22log_{10}{\left(d_{3D}^{(m,n)}\right)} + 20log_{10}{\left(f_{c}\right)}
\end{equation}
\begin{equation} \label{eq4}
\begin{split}
PL_{NLOS}^{(m,n)} = 13.54 + 39.08log_{10}{\left(d_{3D}^{(m,n)}\right)} \\ + 20log_{10}{\left(f_{c}\right)}-0.6	\left(h_{n}-1.5\right) 
\end{split}
\end{equation}
where $d_{3D}^{(m,n)} = \sqrt{(x_m-x_n)^2+(y_m-y_n)^2+(h_m-h_n)^2}$ denotes the distance between the antennas of UAV-BS $m$ and user $n$, while $f_{c}$ is the carrier frequency. Hence the average path loss between UAV-BS $m$ and user $n$ can be denoted as:\\
\begin{equation} \label{eq5}
\begin{split}
PL_{MN}^{(m,n)} = Pr_{LOS}^{(m,n)} \times PL_{LOS}^{(m,n)} + \\ \left(1-Pr_{LOS}^{(m,n)}\right) \times PL_{NLOS}^{(m,n)}
\end{split}
\end{equation}
Similarly, $PL_{DN}^{(d,n)}$ denotes the average path loss between macro-BS $d$ and user $n$, while $PL_{DM}^{(d,m)}$ denotes the average path loss between macro-BS $d$ and UAV-BS $m$. To indicate whether a sub-channel is occupied by a UAV-BS/macro-BS to serve the users, an occupy indicator is defined and setting $c_{i}^{k}$ as 1 implies that the sub-channel $k$ is occupied by UAV-BS/macro-BS $i$. Meanwhile, another indicator is defined where $\alpha_{m}^{n}=1$ indicating that user $n$ is served by UAV-BS $m$. Hence, the SINR between UAV-BS $m$ and user $n$ on sub-channel $k$ can be denoted as:
\begin{equation} \label{eq6}
\begin{split}
\Upsilon_{(n,m)}^{k} = \frac{c_{m}^{k}\alpha_{m}^{n}\left( P_m - PL_{MN}^{(m,n)} \right)}{N_{0}B_k + \sum_{i\neq m} c_{i}^{k}\left( P_i - PL_{I}^{(i,n)} \right) }
\end{split}
\end{equation}
where $P_i$ represents the transmit power of interfering node $i$ and $N_{0}$ is the power spectral density of the additive Gaussian noise. When the UAV-BSs are serving users, the interference may not only come from the other UAV-BSs serving users but also from the macro-BSs serving other UAV-BSs/users. Therefore, $PL_{I}^{(i,j)}$ generally denotes the path loss between user/UAV-BS $j$ and interfering node (UAV-BS/macro-BS) $i$. Based on the above-mentioned expressions, the achieved throughput for MC user can be obtained by:\\
\begin{equation} \label{eq7}
\begin{split}
C_{n} = \sum_k^K \lambda_n Blog_{2}\left( 1 + \Upsilon_{(n,m)}^{k} \right)
\end{split}
\end{equation}
where $\lambda_n$ is the user drop indicator. The user $n$ with an SINR lower than $\Upsilon_{min}$ will be dropped and its corresponding user drop indicator $\lambda_n$ equals zero. For the drop rate of MC users which will be used in the following sections, it is defined as:\\
\begin{equation} \label{eq8}
\begin{split}
\beta_{_{MC}} = \frac{\sum_n^{N_{_{MC}}} \lambda_n}{N_{_{MC}}}, \lambda_n
\end{split}
\end{equation}
where $N_{_{MC}}$ is the number of MC users.

\subsection{Problem Formulation}
In a multi-network scenario consisting of both existing BSs on the ground and temporarily deployed UAV-BS, the deployment of the UAV-BS play a critical role in guaranteeing the performance of the target users/services (e.g., MC users/services). It can also impact the overall system performance. As the UAV-BS is connected to the core network using wireless backhaul, it is important to ensure the good quality of both the backhaul and access links when performing this system optimization. Furthermore, the optimal solution depends on many factors like network traffic load distribution, quality of service (QoS) requirements and user movements on the ground. Therefore, jointly optimizing these parameters of UAV-BS is a complex system-level optimization problem that needs to be solved in a dynamic changing environment.

In order to best serve target users while also maintaining a good backhaul link quality between UAV-BSs and their donor-BSs, we aim to solve the following research problems: 1) Design an RL algorithm to jointly optimize the 3-D locations of the UAV-BSs. 2) Find the movement strategy of a UAV-BS to accommodate the dynamically changing user distribution. 

Based on the system model introduced in the previous sub-section, the target problem we intend to solve is optimizing the 3-D locations of the UAV-BSs to maximize a weighted sum of the following system key performance metrics for the MC users:

\vspace{-1pt}
\begin{itemize}
\item Backhaul link rate for UAV-BS: On one hand, the backhaul link rate reflects the link quality when the UAV-BS is served as a user via its donor-BS. On the other hand, it also affects the end-to-end throughput performance of its associated users since the throughput of UAV-served users is calculated by considering the quality of both the access link and backhaul link. 
\item The 5-percentile and 50-percentile of the cumulative distribution function (CDF) of MC user throughput:  The 5-percentile MC user throughput represents the performance of the cell-edge MC users, i.e., the MC users with the "worst" throughput performance, while the 50\% throughput indicates the average MC user performance in the simulation area. 
\item Drop rate for MC users: The ratio of MC users that cannot be served with the required services. This is an important performance metric for MC scenarios, since for MC users, keeping reliable connectivity broadly is more important than guaranteeing high-demand services for specific users in most MC cases.
\end{itemize}
\vspace{-1pt}

\section{ML-based Solution}

In this section, we describe how we transform and model the considered use case in an ML environment. Three important components, including the state space, action space, and reward function, are constructed in order to design an RL algorithm to jointly optimize the 3-D position of multiple UAV-BSs in an IAB network.

\subsection{Modeling of ML Environment}

\subsubsection{State Space}

In our case, a UAV-BS state at a given time instance $t$ has three dimensions, namely the UAV-BS's 3-D position. 

We use $\mathcal{P}_t = \{x_t, y_t, z_t\}$ to denote the 3-D position of a UAV-BS at time $t$. Then, a UAV-BS's state at a given time instance $t$ is denoted as $s_t=\{x_t, y_t, z_t\}$. Table \ref{tab1:candi} shows the candidate values for each UAV-BS:

\begin{table}[h]
\caption{Candidate values for each UAV-BS in the simulation environment}
\label{tab1:candi}
\setlength{\tabcolsep}{3pt}

\begin{tabular}{m{75pt}<{\centering} m{160pt}<{\centering}}
\toprule
3-D position $\mathcal{P}$ Space & UAV1 Candidate Values\\
\midrule
$x$ & $[85, 257, 428, 600]$ meters
 \\

$y$ & $[-514, -342, -171, 0]$ meters
 \\
 $z$ & $[10, 20]$ meters
 \\
\bottomrule
\toprule
3-D position $\mathcal{P}$ Space & UAV2 Candidate Values\\
\midrule
$x$ & $[-600, -428, -257, -85]$ meters
 \\

$y$ & $[-514, -342, -171, 0]$ meters
 \\
 $z$ & $[10, 20]$ meters
 \\
\bottomrule

\toprule
3-D position $\mathcal{P}$ Space & UAV3 Candidate Values\\
\midrule
$x$ & $[-428, -257, -85, 85, 257, 428]$ meters
 \\

$y$ & $[171, 342, 514]$ meters
 \\
 $z$ & $[10, 20]$ meters
 \\
\bottomrule
\end{tabular}

\end{table}
It should be noted in Table \ref{tab1:candi} that the available height range for all UAV-BSs is limited between 10 m and 20 m, rather than deploying the UAV-BSs into a higher altitude. The reason is that, in the scenario considered in this paper, the UAV-BSs tend to stay at a lower height to maintain good backhaul links to on-ground donor-BS and also provide better access links to serve on-ground MC users, which makes the current height range selection reasonable.

The candidate values of 2-D space location $x$ and $y$ axis cover the disaster area shown in Figure \ref{fig:SimulationScenario}. The location options are selected by three deployed UAV-BSs. The 2-D MC area has been divided into 3 parts, with each UAV-BS covering one part of the area. For UAV1, $x$ and $y$ axis options are $[85, 257, 428, 600]$ and $[-514, -342, -171, 0]$ meters. For UAV2, $x$ and $y$ axis options are $[-600, -428, -257, -85]$ and $[-514, -342, -171, 0]$ meters. For UAV3, $x$ and $y$ axis options are $[-428, -257, -85, 85, 257, 428]$ and $[171, 342, 514]$ meters. And the height options for all three UAV-BSs in the $z$ axis are $[10, 20]$ meters. As a result, the total number of state combinations in this environment is 18928. The computation complexity will be linearly increased $O(n)$ based on the total number of input states combination.

\subsubsection{Action Space}

In order to enable a UAV-BS to control its state, for each state dimension,  we defined three potential action options and the UAV-BSs choose an action from three candidate options. These three alternative action options are denoted by the three digits: $-1, 0, 1$, where ``-1" indicates that a UAV-BS decreases the status value at this state dimension by one step from its current value; ``0" indicates that a UAV-BS does not need to take any action at this state dimension and keeps its current value; "1" indicates that a UAV-BS increases the status value at this state dimension by one step from its current value.

For example, if the x-axis value of the UAV1 (i.e. the value of the $x_t$ dimension) equals 257 meters, an action coded by ``-1" for this dimension means that the UAV-BS will select an action to reduce position value to 85 meters, an action coded by ``0" implies that the UAV-BS will hold the current position ($257$ meters), and an action coded by ``1" implies that the UAV-BS will increase the position value to $428$ meters. The same policy is applied to all dimensions of the state space.

\begin{figure}[!htbp]
\centerline{\includegraphics[scale=0.35]{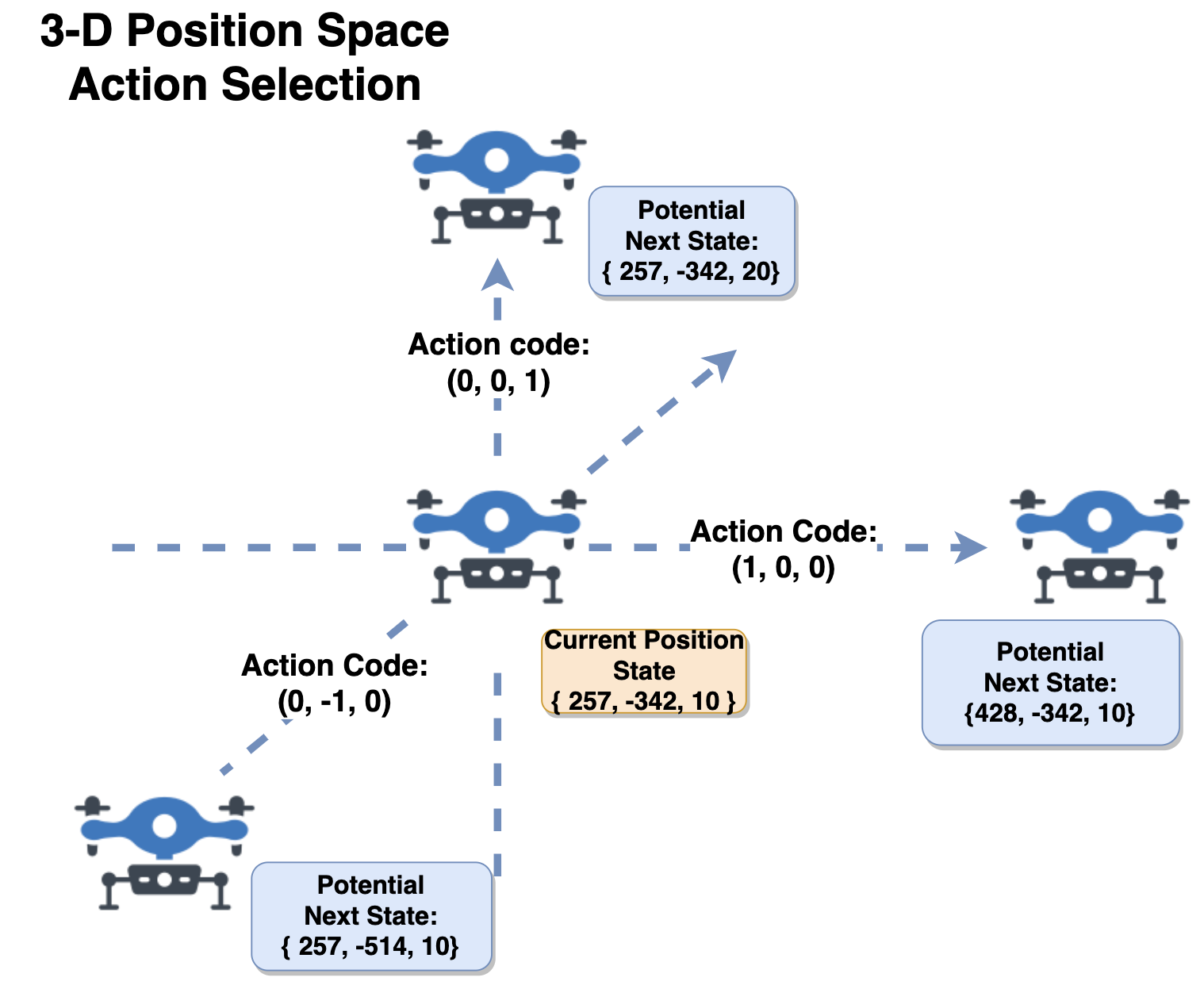}}
\caption{Example of the UAV1's state transition from current state $\{ 257, -342, 10\}$ meters}
\vspace{-10pt}
\label{fig:action_selection}
\end{figure}

Since there are three action alternatives for each space state, the action pool for 3-D position space, the pool has 27 action candidates that may be programmed to an action list $\mathscr{P}$ =[(-1, -1, -1), (-1, -1, 0), (-1, -1, 1), (-1, 0, -1) ..., (1, 1, 1)]. As a result, if we combine the action of the 3-D position space, at any given moment $t$, a UAV-BS can thus choose an action $a_t$ from these 27 alternatives. Figure \ref{fig:action_selection} depicts a state transition from the specified state $s_t=\{257, -342, 10\}$ meters.

\subsubsection{Reward Function Design}

It is more critical to serve as many MC users as possible with appropriate service quality than to maximize the peak rate of a subset of MC users. As a result, the aggregated reward metrics are produced for the reward function design of the reinforcement learning algorithm to take into account both the impact of other drones' actions as well as the quality of local services. Therefore, the reward is calculated using the average of the performance indicators of local and neighbouring agents. We have selected six key performance metrics for each local agent to highlight the local quality of service for MC users, including:

\vspace{-1pt}
\begin{itemize}

\item The drop rates of MC users for 
UL and DL ($\beta_{ul}, \beta_{dl}$), which reflect the percentage of unserved MC users.

\item The 50\% throughput values of MC users for both UL and DL ($\alpha_{ul-50\%},  \alpha_{dl-50\%}$), which represent the average performance of the MC users, and

\item The 5\% throughput values of MC users for both UL and DL ($\alpha_{ul-5\%}, \alpha_{dl-5\%}$), which represent the ``worst" performance of the MC users.

\end{itemize}
\vspace{-1pt}

The reward function is built as a weighted sum of these six feature values to balance these critical performance indicators, as shown below. The reason why backhaul link rate is not considered here is that the values of the six features all rely on the quality of backhaul link between the UAV-BS and its donor-BS. Before the model, all characteristics are normalised using min-max normalization, thus the values are constrained within the range $[0, 1]$.

\begin{equation} \label{eq9}
\begin{split}
R_s = & \omega_1 \times \frac{(1 - \beta_{dl}) + (1 - \beta_{ul})}{2} + \omega_2 \times \frac{(\alpha_{ul-5\%} + \alpha_{dl-5\%})}{2}\\
& + \omega_3 \times \frac{(\alpha_{ul-50\%} +  \alpha_{dl-50\%})}{2}
\end{split}
\end{equation}

Furthermore, we set $ \omega_1 + \omega_2 + \omega_3 = 1$ to normalize the reward value such that $R_s$ is between $[0, 1]$. To emphasize the significance of supporting all MC users, we assign higher weights to user drop rates and 5\% MC-user throughput metrics. This is because, in the MC use cases, we must first prioritize that all users have access to the communication service rather than focusing on optimizing the communication quality of a small subset. In this paper, our method uses the weight values $\omega_1 =0.5$, $\omega_2=0.3$ and $\omega_3=0.2$. In order to know the influence of each UAV-BS, the reward function will also aggregate the reward values of the neighbour agents. Hence, the following is the reward function which is applied in the algorithm:

\begin{equation}
\begin{aligned}
R_s = \frac{\sum_{c=1}^{C}\mathcal{M}^c + \mathcal{M}}{len(C) + 1}\\
\end{aligned}
\end{equation}

Assuming that $C$ is the set of register neighbours, $\mathcal{M}$ represents the current agent's local system performance and the $\mathcal{M}^c$ indicates the local system performance of its neighbour ID $c$.

\subsection{RL Algorithm Design}

In this section, we design an RL algorithm to solve the optimization problem of the considered use case. RL is distinct from supervised and unsupervised learning in the field of ML in that supervised learning is performed from a training set with annotations provided by an external supervisor (task-driven), whereas unsupervised learning is typically a process of discovering the implicit structure in unannotated data (data-driven). RL is suitable for this case since the method provides a unique feature: the trade-off between exploration and exploitation, in which an intelligence agent must benefit from prior experience while still subjecting itself to trial and error, allowing for a larger action selection space in the future (i.e., learning from mistakes).

In order to achieve better self-control decisions for our scenario, we applied deep Q-network (DQN) as our base RL algorithm. The algorithm was first proposed by Mnih et al. in \cite{mnih2013playing}\cite{mnih2015human}  by combining convolutional neural networks with Q-learning algorithms \cite{watkins1989learning} in traditional RL. The approach has been frequently used in gaming and static environments. However, the original approach is incapable of adapting to our MC situation due to environmental changes. To address these issues, we have proposed two significant schemes in our autonomous UAV-BSs control algorithm (Algorithm \ref{alg:DQN}): adaptive exploration control and value-based action selection.

\begin{algorithm}[]

    \caption{Deep Reinforcement Learning in each UAV-BS with adaptive exploration and value-based action selection}
    \label{alg:DQN}
    
    \begin{algorithmic}
	\STATE Initialize the agent's replay memory Buffer $\mathcal{D}$ to \newline capacity $N$
	\STATE Initialize action-value function $Q$ with two random sets \newline of weights $\theta, \theta'$
	\STATE Initialize exploration probability $\varepsilon$ to 1
	
	\STATE Set previous reward value $r_{p}$ to 0

	\FOR{$Iteration = 1,M$}
	    \FOR{$t = 1,T$}
	    
	    \STATE $\mathscr{T}_t, \mathscr{P}_t \leftarrow$  Action\_Selection($r_{p}$, $r_{t}$, $\varepsilon$)
		\STATE $a_t = \{\mathscr{T}_t, \mathscr{P}_t\}$
		\STATE Set $r_p = r_t$
		\STATE Decode $a_t$ to action options in four state \newline dimensions and execute the actions
		\STATE Collect reward $r_{t}$ and observe the agent's next state 
		\STATE $\mathcal{P}_{t+1} \leftarrow \{x_{t+1}, y_{t+1}, z_{t+1}\}$
		\STATE Set $s_{t+1} = \{\mathcal{T}_{t+1}, \mathcal{P}_{t+1}\}$
		\STATE Store the state transition $(s_t, a_t, r_{t}, s_{t+1})$ in $\mathcal{D}$
		
		$\mathcal{A}_s$, $\mathcal{A}_o \leftarrow$ Action\_Grouping($a_t$)
		
		\STATE Sample mini-batch of transitions $(s_j, a_j, r_{j}, s_{j+1})$ from buffer $\mathcal{D}$
		\IF {$s_{j+1}$ is terminal}
		    \STATE Set $ y_j =  r_{j}$  
		\ELSE
		    \STATE Set $y_j = r_{j} + \gamma \max_{a'} Q(s_{j+1}, a'; \theta')$
		\ENDIF
		\STATE Perform a gradient descent step using targets $y_j$ \newline with respect to the online parameters $\theta$ 
		\STATE Set $\theta' \leftarrow \theta$
		\STATE $\varepsilon \leftarrow $ Adaptive\_Exploration($r_{p}$, $r_{t}$, $\varepsilon$) 
	    \ENDFOR
	\ENDFOR
	
    \end{algorithmic}
   
\end{algorithm}

\setlength{\textfloatsep}{0.001cm}
\setlength{\floatsep}{0.001cm}

\subsubsection{Adaptive Exploration (AE)}

Because of the environmental changes, the original DQN model needs to be updated to accommodate feature value changes. As a result, we create a dynamic exploration probability triggered by a substantial decline in reward value. Following the completion of each learning iteration, the final reward value is checked and compared to the pre-defined reward drop and upper reward thresholds. Based on the outcome, the exploration probability $\epsilon$ will be adjusted.

Each UAV-BS initially explores the state space and then performs Q-value iterations at each training episode. When deciding whether to take an action that gives the maximum reward value or randomly explore a new state, a $\epsilon$-greedy exploration is used. The parameter $\epsilon$ determines the likelihood of exploration. Each training step's data is saved in a replay batch $\mathcal{D}$. Each row of $\mathcal{D}$ holds the tuple $(s_t, a_t, r_{t}, s_{t+1})$, which represents the current state, action, reward, and next state for a training step. Samples will be chosen at random and used to update the Q value model.

The most recent reward value is reviewed and compared to a pre-defined reward-drop threshold and an upper reward threshold. Then, the exploration probability is updated by checking the following three conditions: (Algorithm \ref{alg:AE}):

\begin{itemize}

\item If the most recent reward value is less than the prior reward, and the difference is greater than the reward drop threshold, the exploration probability is increased to 0.1.

\item If the most recent reward value exceeds the higher reward threshold, we can conclude that the algorithm has already located the optimal zone capable of delivering a reliable connection to MC users. The likelihood of exploration will be matched to the probability of completion.

\item Otherwise, the exploration probability will multiply by an exploration decay and fall linearly after each learning cycle.

\end{itemize}

\begin{algorithm}[h]

    \caption{Adaptive Exploration Algorithm (AE)}
    \label{alg:AE}
    
    \begin{algorithmic}

    \STATE \textbf{Set} restarting exploration probability $\varepsilon_{Restart}$ \newline to 0.1
	\STATE \textbf{Set} ending exploration probability $\varepsilon_{End}$ to 0.0001
	\STATE \textbf{Set} exploration decay $\varrho$ to 0.995
	
	\SetKwFunction{FMain}{Adaptive\_Exploration} 
    \SetKwProg{Fn}{Function}{:}{}  
    \Fn{\FMain{$r_{p}$, $r_{t}$, $\varepsilon$}}{
        \IF {$r_{p}$ - $r_t$ $>$ Drop threshold}
	    \STATE Set $\varepsilon$ = $\varepsilon_{Restart}$
	    \ELSIF {$r_t$ $>$ Upper reward threshold}
	    \STATE Set $\varepsilon = \varepsilon_{End}$
	    \ELSE
	    \STATE Set $\varepsilon = \varepsilon \times \varrho$
	    \ENDIF
	    
        \textbf{return} $\varepsilon$
        
}

    \end{algorithmic}
   
\end{algorithm}

\begin{algorithm}[h]

    \caption{Value-based action selection (VAS)}
    \label{alg:vas}
    
    \begin{algorithmic}

	\STATE \textbf{Set} grouping threshold $\beta$ to 0
	
	\SetKwFunction{FMain}{Action\_Grouping} 
    \SetKwProg{Fn}{Function}{:}{}  
    \Fn{\FMain{$a_t$}}{
        \STATE $\mathscr{P}_t \leftarrow a_t = {\mathscr{T}_t, \mathscr{P}_t}$ 
        \FOR {all potential next action $\mathscr{P}_{t+1}$}
		\IF {$\vec{\mathscr{P}}_t \cdot \vec{\mathscr{P}}_{t+1} > \beta$}
		    \STATE Append $\vec{\mathscr{P}}_{t+1}$ to $\mathcal{A}_s$ 
		\ELSE
		    \STATE Append $\vec{\mathscr{P}}_{t+1}$ to $\mathcal{A}_o$
		\ENDIF
		\ENDFOR

        \textbf{return} $\mathcal{A}_s$, $\mathcal{A}_o$
        
    }
    \STATE
    
    \vspace{10pt}
    \SetKwFunction{FMain}{Action\_Selection} 
    \SetKwProg{Fn}{Function}{:}{}  
    \Fn{\FMain{$r_{p}$, $r_{t}$, $\varepsilon$}}{
        \IF {$r_t \geq r_p$}
		    \STATE Select a random action $\mathscr{P}_t$ with probability $\varepsilon$ \newline from the same consequence 3-D position \newline action pool $\mathcal{A}_s$ 
		   
		\ELSE
		    \STATE Select a random action $\mathscr{P}_t$ with probability $\varepsilon$ \newline from the opposite consequence 3-D position \newline action pool $\mathcal{A}_o$ 
		    
		\ENDIF
		\STATE Otherwise, select $a_t = {\arg\max}_a Q(s_t, a; \theta)$
	    
        \textbf{return} $\mathscr{T}_t, \mathscr{P}_t$
        
    }
    
    \end{algorithmic}
   
\end{algorithm}

\setlength{\textfloatsep}{0.001cm}
\setlength{\floatsep}{0.001cm}

\begin{figure}[h]
\centerline{\includegraphics[width=3.5in]{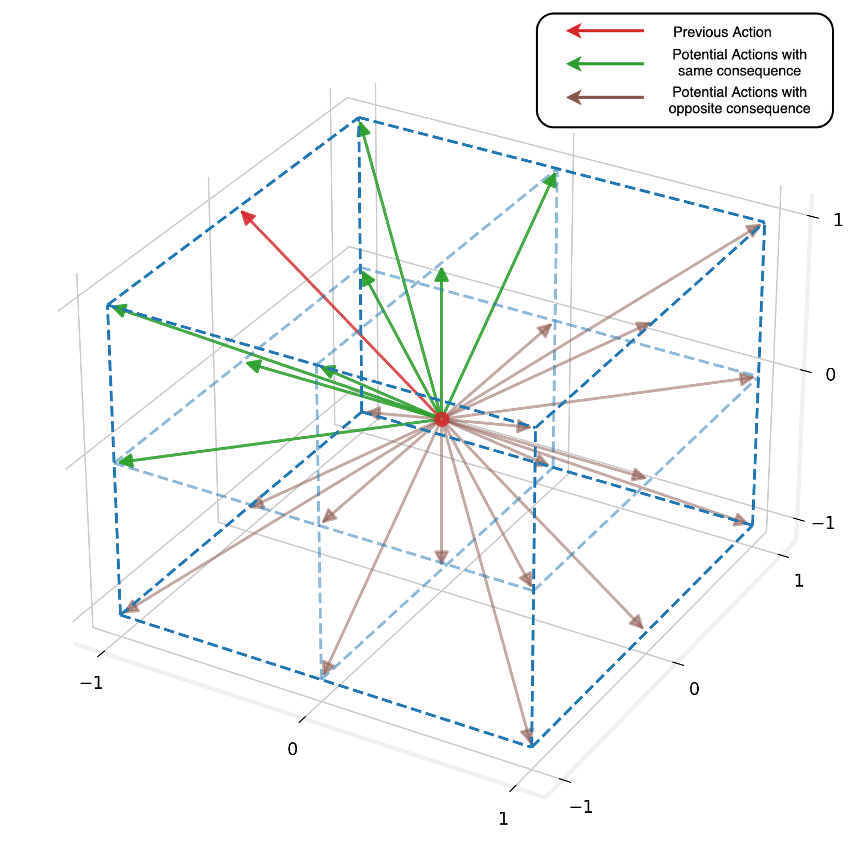}}
\caption{Diagram of a potential set of next actions with same or opposite consequence}
\label{fig:next_action}
\end{figure}

For the hyper-parameters of the deep Q-network, we explored various sets of combinations in order to achieve acceptable results. During the training, the discount factor is set to 0.99, the learning rate is set to $5 \times 10^{-5}$ and the number of training iterations equals 1000.

\subsubsection{Value-Based Action Selection (VAS)}

Although the $\epsilon$-greedy algorithm can strike a reasonable balance between exploration and exploitation, in some cases the approach utilized for exploration is redundant and time-consuming. The algorithm will choose actions at random throughout the searching stage, which may lengthen the search time.
However, when dealing with a large action and state space, random action selection is clearly not an effective strategy and may cause decision-making to be delayed, which is unacceptable in most time-critical businesses. Therefore, we propose a novel value-based action selection strategy (Algorithm \ref{alg:vas}) which can lead to fast decision-making for a UAV-BS when determining its 3-D space location. 

As described in the previous section, an agent's 3-D position state at a given time instance $t$ is denoted as $\mathcal{P}_t=\{x_t,y_t,z_t\}$. Since each position state has three dimensions and each state dimension has three action options, the action pool contains in total 27 action candidates that can be programmed to a list of action space $[(-1, -1, -1), (-1, -1, 0), (-1, -1, 1),... (1, 1, 1)]$. Each element in this list can then be regarded as an action vector.

Figure \ref{fig:next_action} depicts a probable set of next actions with the same or opposite consequence. The consequence is defined as the reward value (or monitored performance metrics) change after an action has been executed. The algorithm will analyze the outcome of past actions. If the prior action decision has a positive outcome (the reward value increases or monitored performance metrics become better) as defined above, the algorithm will choose actions from a pool of following actions with the same consequence. The dot product between two action vectors determines the result. If the dot product is larger than 0, this action vector can be assumed to have the same outcome as the prior action option.

If the previous action decision results in a negative consequence (the reward value decreases or monitored performance metrics become worse), the algorithm will select actions from the pool consisting of potential next actions with the opposite consequence. The opposite consequence is determined by the dot product of two action vectors that is smaller than or equal to 0. The actions in this pool will result in an opposite consequence compared with the previous action decision. Assume that the previous action vector is $\vec{\mathscr{P}}_t$ while the next potential action vector is $\vec{\mathscr{P}}_{t+1}$:

\begin{equation}
    \begin{cases}
    \vec{\mathscr{P}}_t \cdot \vec{\mathscr{P}}_{t+1} > 0       & \quad \text{Same consequence as previous } \\
    \vec{\mathscr{P}}_t \cdot \vec{\mathscr{P}}_{t+1} \leq 0 & \quad \text{Opposite consequence as previous} 
  \end{cases}
\end{equation}

In Figure \ref{fig:next_action}, the red vector represents the previous action decision. The angle between the previous action vector (red vector) and the green vectors is less than $\frac{\pi}{2}$, which can be represented by a dot product greater than zero. As a result, the green vectors represent actions that may result in the same consequence as the red vector. Similarly, the angle between the previous action vector (red vector) and the brown vectors is greater than or equal to $\frac{\pi}{2}$, which is represented by a dot product value less than 0. As a result, the brown vectors may have the opposite consequence.

During the UAV-BSs deployment, the algorithm monitors a set of critical system performance values (the reward value). Based on the current and a set of previous performance values, the algorithm will evaluate the consequences caused by the previous action. The algorithm will thus select the action set which will potentially result in positive consequences.


\subsection{Decentralized Reinforcement Learning}

\begin{figure}[h]
\centerline{\includegraphics[width=3.2in]{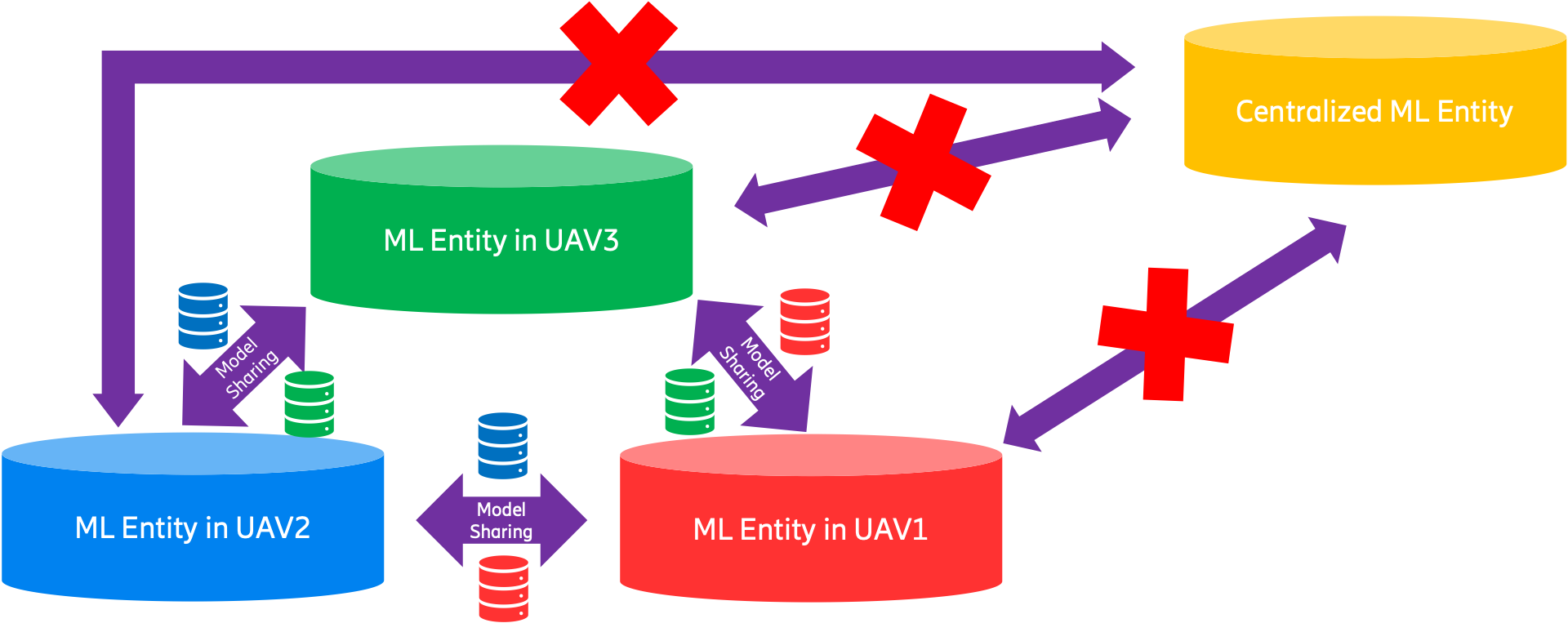}}
\caption{Decentralized Architecture for Multi-UAV Coordination}
\label{fig:deentity}
\end{figure}

In some circumstances, a single UAV is not capable to be extended to cover a larger area. As shown in Figure \ref{fig:MCScenario}, multiple UAV-BSs are deployed to work together to service the MC users. An extensible decentralized method for deploying numerous UAV-BSs is therefore designed. The concept is illustrated in Figure \ref{fig:deentity} where we relocate the central server operation function from the central entity and attach it to the edge entity on UAV, as opposed to the typical single-agent reinforcement learning algorithms, to achieve decentralized characteristics.

\begin{algorithm}[h]

    \caption{Transmission functions of the decentralized reinforcement learning algorithm (DecRL)}
    \label{alg:dec-DQN}
    
    \begin{algorithmic}

	\FOR{$Iteration = 1,M$}
	    \FOR{$t = 1,T$}
            \STATE After action selections:
		\FOR{each client $c \in C$ \textbf{in parallel}} 
		
		 \STATE send $\{s_{t+1}, \mathcal{M}\}$; 
		 \STATE receive $\{s_{t+1}^c, \mathcal{M}^c\}$
             
        \ENDFOR
		$s_{t+1} = \{ (x_{t+1}, y_{t+1})^{c}$ $for$ $c$ $in$ $C \}$
		\STATE Store the state transition $(s_t, a_t, r_{t}, s_{t+1})$ in $\mathcal{D}$
		\STATE Learning in each UAV-BS: $f_{AE\&VAS}(s_t, a_t, r_{t}, s_{t+1})$
		
		\ENDFOR
		
		\IF{$t$ mod $f$ == 0}
            
            \FOR{each client $c \in C$ \textbf{in parallel}} 
            \STATE send $\theta'_{t+1}$; 
            \STATE receive ${\theta'_{t+1}}^c$; 
            \ENDFOR

        \STATE $ \theta'_{t+1} \longleftarrow \sum_{c=1}^{C}\frac{1}{len(C)}{\theta'_{t+1}}^c;$ 
            
		\ENDIF

	\ENDFOR

\textbf{End Function}

\end{algorithmic}
   
\end{algorithm}

The system has two different kinds of data for exchanging information, namely the system-related data (including location information and system KPIs) and the model data. The location  will communicate with nearby drones regarding the connection performance and UAV-BS system-related data. These kinds of data can assist each drone in understanding how their movements affect the others and in being aware of one another's surroundings. Following each UAV-BS decision, the information will be continuously exchanged and used as a guide for the subsequent choice. The local model of each UAV-BS will be shared with its neighbours via the model data channel, which is indicated by the green line. Each UAV-BS has a separate procedure to train, communicate, and receive model weights and service metrics during the learning process. Each UAV-BS will share their learning experiences as a result, and the others can gain information from the experiences of the others. After multiple training epochs, the UAV-BSs can swap their model with their neighbours under the control of a frequency parameter. The process is described in Algorithm \ref{alg:DQN}. The procedures can be summarized as follows:

\begin{enumerate}[Step 1:]

\item Each training episode will begin with each UAV-BS exploring and locating its neighbours before moving on to exploring the environment and doing Q-value iterations. When deciding whether to choose the best action or to randomly explore the new state, a $\epsilon$-greedy exploration is used. The parameter $\epsilon$ specifies the likelihood of exploration.

\end{enumerate}

\begin{enumerate}[Step 2:]

\item After making a choice, each UAV-BS will notify its neighbours of the state and local performance indicators. The agent will simultaneously listen to the other neighbours, and get ready to receive their states $s_{t+1}^c$ and local performance metrics $\mathcal{M}^c$. A global system metric value that can direct each UAV-BS to take future actions will be formed when all metrics have been received and the reward has been calculated based on the reward function $R_s$.

\end{enumerate}

\begin{enumerate}[Step 3:]

\item  A replay batch $\mathcal{D}$ contains the data for each training stage. The tuple $(s_t, a_t, r_{t}, s_{t+1})$, or the current state, action, reward, and next state for a training step, is contained in each row of $\mathcal{D}$. For the purpose of updating the Q value model, samples will be chosen at random.
The current state reward pairs will also be distributed to the other agents after each decision round.

\end{enumerate}

\begin{enumerate}[Step 4:]

\item  A UAV-BS will send the updated model results, $\theta'$, to its registered neighbours for model aggregation after it has reached the predetermined exchanging iteration. Each UAV-BS will simultaneously listen to its neighbours in order to receive models and service metrics.
\end{enumerate}

\begin{enumerate}[Step 5:]
\item Each node executes aggregation by averaging all updated models depending on the aggregation function, $ \theta'_{t+1} \longleftarrow \sum_{c=1}^{C}\frac{1}{len(C)}{\theta'_{t+1}}^c;$,  after receiving all the models from the registered neighbours. 
\end{enumerate}

\begin{enumerate}[Step 6:]
\item The updated model is used by the edge device to replace the outdated one and to carry out additional local training. We'll repeat the steps from above.

\end{enumerate}

\section{Simulation Results and Analysis}
In this section, the simulation configuration and scenario deployment are introduced firstly. Then, we investigate the impact of the 3-D location of multiple UAV-BSs on the performance of MC users in terms of backhaul link rate, throughput, and drop rate based on system-level simulations. Finally, we present the results of proposed RL algorithms for autonomous UAV-BS navigation.

\subsection{Simulation Configuration}
To evaluate the performance of the proposed RL algorithm in solving the formulated problem, we build a multi-cell scenario by considering the predefined system model, and a simulation is executed with a system-level simulator. With the output of the simulation, the proposed RL algorithm can be applied for UAV-BS to build a well-trained model, based on which the optimal UAV-BS position and antenna configuration can be found rapidly.
\begin{figure}[!t]
  \begin{center}

    \includegraphics[scale=0.65]{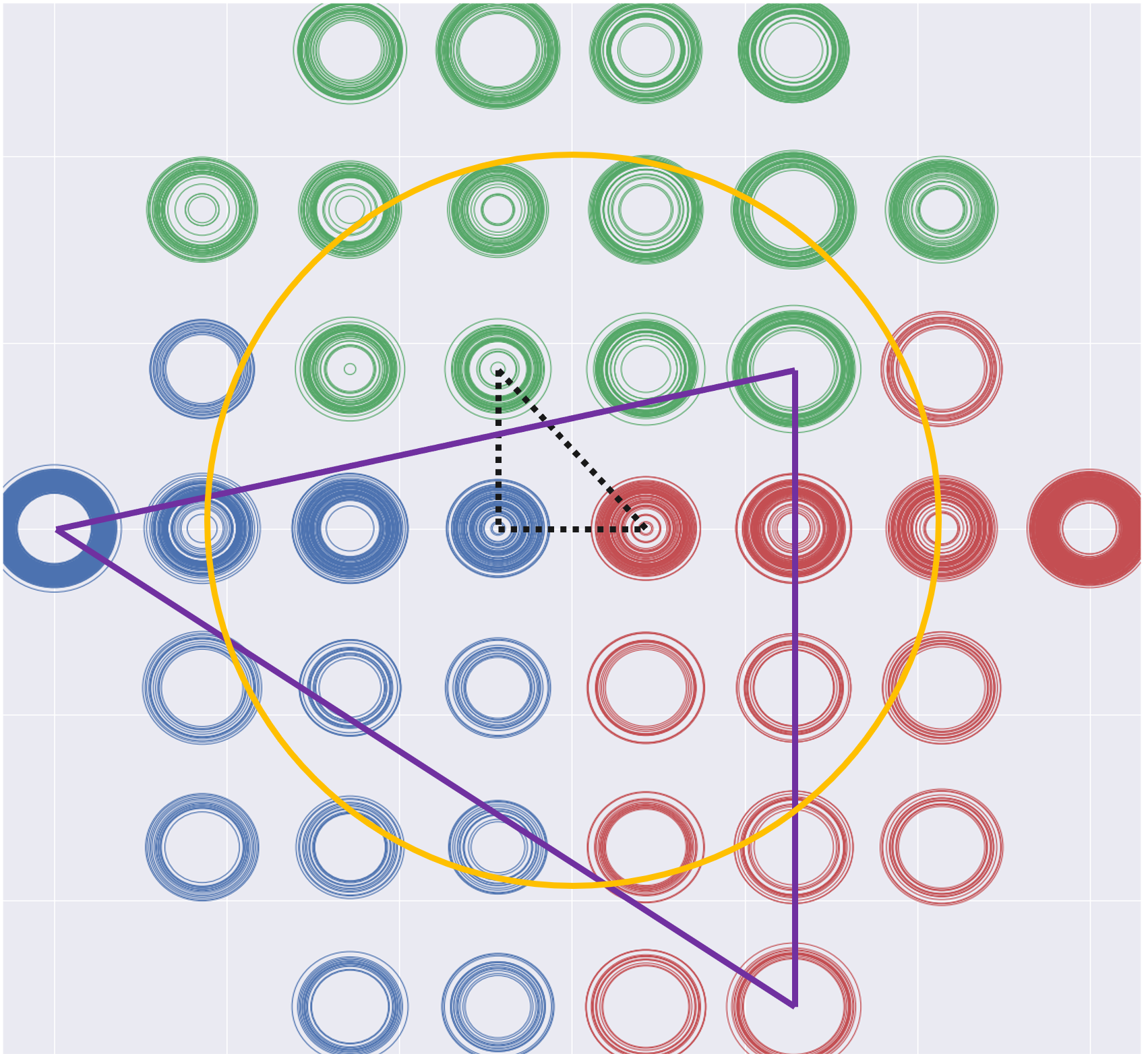}

    \caption{Impact of UAV-BSs' Positions on average backhaul link rate}

    \label{Fig:UAVPositionImpact}
    
  \end{center}

\end{figure}

In the simulation, we drop 500 users in the area as shown in Figure \ref{fig:SimulationScenario}. The circle area with a 350m radius around the UAV-BS is defined as the MC area. The users located in the MC area are marked as MC users, while the others are normal users. All users follow an arrival model and only arrived users can be considered as activated. 
To investigate how a well-trained RL model performs in a dynamic environment, we design a set of different user distributions to simulate the case of slow-moving users.


The detailed simulation parameters are shown in Table \ref{Table:SimulationParameters}.

\begin{table}[h]
\caption{Simulation Parameters}
\label{Table:SimulationParameters}
\setlength{\tabcolsep}{3pt}

\begin{tabular}{|m{110pt}<{\centering}|m{130pt}<{\centering}|}
\hline
Parameter& 
Value\\
\hline
Carrier Frequency& 
3.5 GHz \\
Bandwidth& 
100 MHz \\
Duplex Mode& 
TDD \\
TDD DL/UL Configuration&
DDUD\\
Inter-Site-Distance (ISD)&
750-1000 m\\
Radius of MC Area& 
350 m \\
Number of BSs&
Macro-BS: 6; UAV-BS: 1\\
BS Transmit Power& 
Macro-BS: 46 dBm; UAV-BS: 40 dBm\\
Noise Figure& 
7 dB\\
BS Height&
Macro-BS: 32 m; UAV-BS: 10-300 m\\
Number of Sectors per Site&
3\\
Number of MC\&Normal Users& 
500 \\
User Arriving Rate per Simulation Area& 
\makecell[c]{270 users/s}\\
User Speed& 
3 km/s \\
Minimum Distance between BS and Users& 
30 m \\
Simulation Time& 
2 s \\
\hline
\end{tabular}
\vspace{10pt}
\end{table}

\begin{figure*}[!t]
\centering

\includegraphics[width=\textwidth]{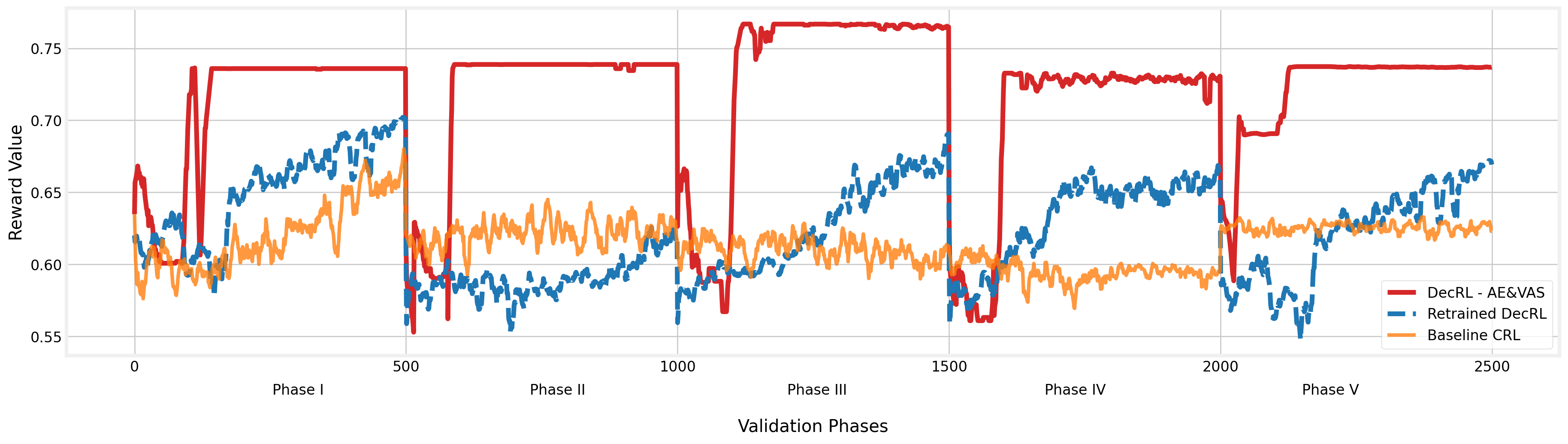}

\caption{Reward value with the number of learning iterations in five consecutive validation phases}
\vspace{-10pt}
\label{fig:Reward}
\end{figure*}

\subsection{System-level Performance Evaluation}
To evaluate the impact of UAV-BSs' positions on the user performance, Figure \ref{Fig:UAVPositionImpact} shows the range of achieved average backhaul link rate when the three UAV-BSs are deployed at different candidate positions as indicated in Figure \ref{fig:SimulationScenario}. Around the MC area, 40 candidate 2D-positions of UAV-BS (colored stars shown in Figure \ref{Fig:UAVPositionImpact}) are selected and indicated by the centers of the circles in Figure \ref{Fig:UAVPositionImpact}. Each colored UAV-BS can only be deployed at the position with the same color. The radius of the circles denotes the normalized average backhaul link rate of three UAV-BSs. For instance, when one UAV-BS is deployed at one candidate position in blue, the other two UAV-BSs can be placed at the all possible combination of green and red candidate positions. This explains why there are multiple circles at each candidate's UAV position. Hence the range of the circle radius at one candidate position denotes the lower and higher bounds of the backhaul link rate when the UAV-BS is placed at the current location. After connecting the locations with the lowest or highest backhaul link rate, two triangles can be observed, where the black dashed circle indicates the positions of three UAV-BSs with the lowest average backhaul link rate, while the solid purple triangle indicates the positions of three UAV-BSs with the highest average backhaul link rate. Based on the distribution pattern of three UAV-BSs, it seems that the optimal UAV-BSs' positions to maximize the backhaul link rate tend to be near the edge of the MC area. This is because the UAV-BSs can keep good backhaul link quality when located near the donor-BSs. 

Based on the above assumptions, the optimal UAV-BSs' positions are different for different performance metrics considered in the reward function. Hence, optimizing the performance metrics in the reward function to achieve the global optimization by adjusting UAV-BSs' positions is a complicated problem, for which ML-based solutions can be applied to find the implicit structure from the collected data.

\subsection{Machine Learning Performance Evaluation}

In this section, we present the experimental results of Decentralized Reinforcement Learning with Adaptive Exploration and Value-based Action Selection (DecRL-AE\&VAS) for autonomously controlling multiple UAV-BSs. For the UAV-BSs deployment, different from the single UAV case introduced in the previous section, more candidate positions are allocated around the MC area for the multi-UAV case. As shown in Figure \ref{fig:SimulationScenario}, the available positions for each UAV-BS don't overlap with others, which means each UAV-BS is covering a certain geographical area with a total of 18928 combinations.  The result is evaluated using two criteria: (1) six features (specified in Section IV - Modeling of ML Environment) that demonstrate link service quality, and (2) model learning quality in each phase as demonstrated by system reward value. The results are compared to several baseline models. As we described before, the experiment contains five validation phases incorporating MC user mobility. When entering a new phase, the algorithms will use the data collected during the new phase to learn and update themselves. During the simulation, the DecRL-AE\&VAS method trains the model from scratch in the training phase and then continuously improves itself in the succeeding validation phases. The previously learnt experience will not be removed for the subsequent sessions. 

There are two baseline methods used to compare the performance of DecRL-AE\&VAS and demonstrate the effectiveness of the decentralized architecture and the convergent efficiency in the training phase, namely, baseline CRL and baseline IRL method. For the baseline IRL algorithm, these baseline models are explicitly trained using each edge UAV-BS. There will not be any model or information exchange between the edge and central nodes during training, in contrast to decentralized reinforcement learning. The service quality performance can be compared to the decentralized reinforcement Learning model to demonstrate how it can outperform those locally trained individual models. For the baseline CRL,  this baseline model is trained using the centralized learning strategy. Prior to model training, all data from the edge are collected into a single server and learn the action strategy step by step.

In the validation phase, in order to prove the performance in a dynamic environment, two baselines are utilized, namely, the retrained DecRL and baseline CRL. The retrained DecRL algorithm removes the past information and randomizes the ML model parameters but with the same decentralized setup. When entering a new phase, the algorithm will retrain the model from scratch. The difference between retrained DecRL and DecRL-AE\&VAS is the utilization of AE and VAS strategies. These strategies have been proven to be useful when deploying UAV-BSs into dynamic environments.  Last but not the least, the baseline CRL algorithm in the validation phases will constantly learn and employ the new data when entering the new phase but with a centralized algorithm which controls all deployed UAV-BSs but without improved strategies.

For the hyper-parameters of our method, we explored various sets of combinations in order to achieve acceptable results. During the training, the network architecture for a DQN is set to $[4-16-16-81]$, the exploration probability decay is set to 0.995, the learning rate is set to $5 \times 10^{-5}$ and the number of learning iterations at the training phase and validation phases equals 1500 and 500.

\begin{figure}[!htpb]
\centering

\includegraphics[scale=0.42]{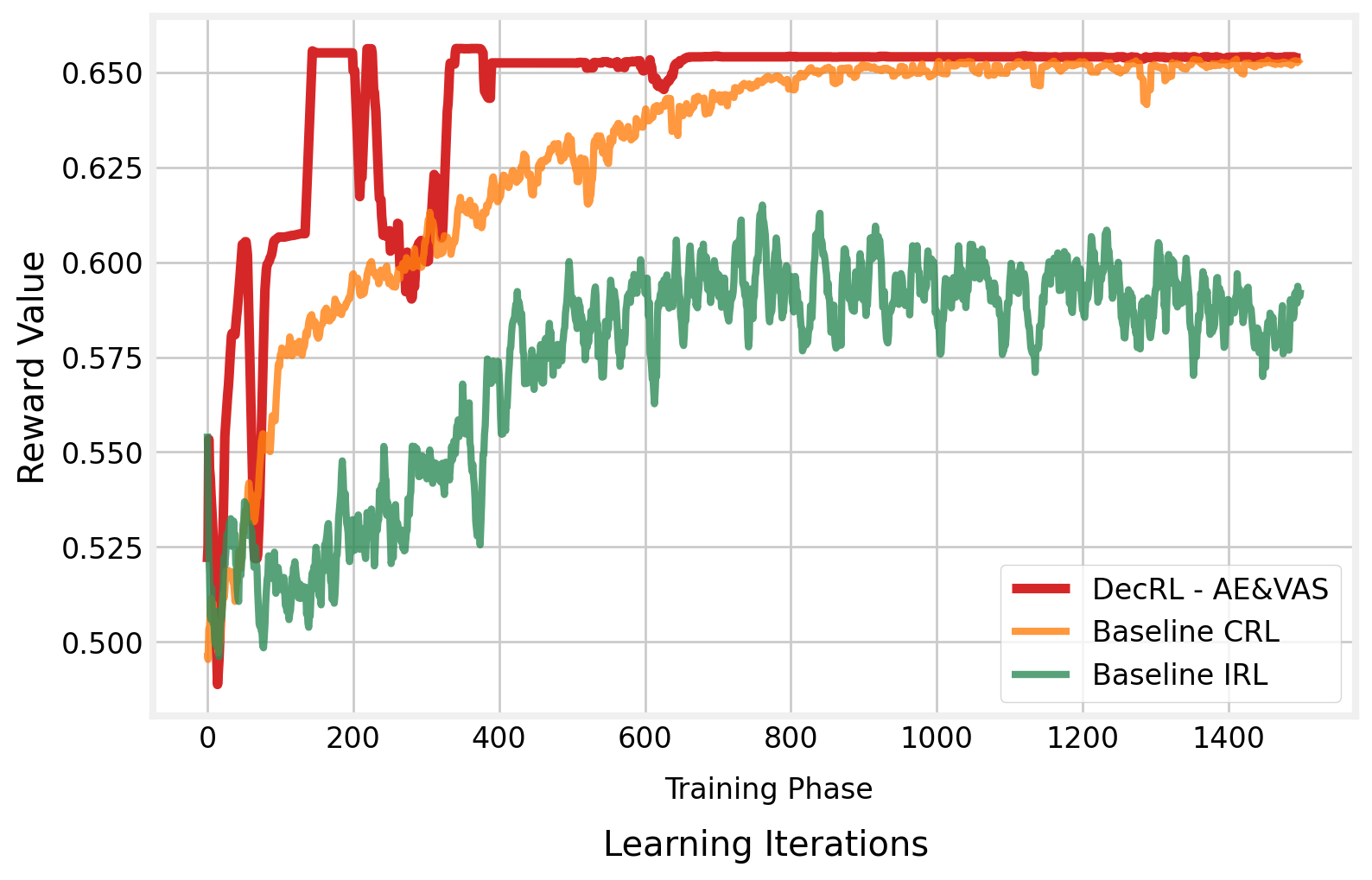}

\caption{Reward value with the number of learning iterations in the training phase}

\label{fig:rewardt}
\end{figure}

\begin{figure}[!htpb]
\centering

\includegraphics[scale=0.35]{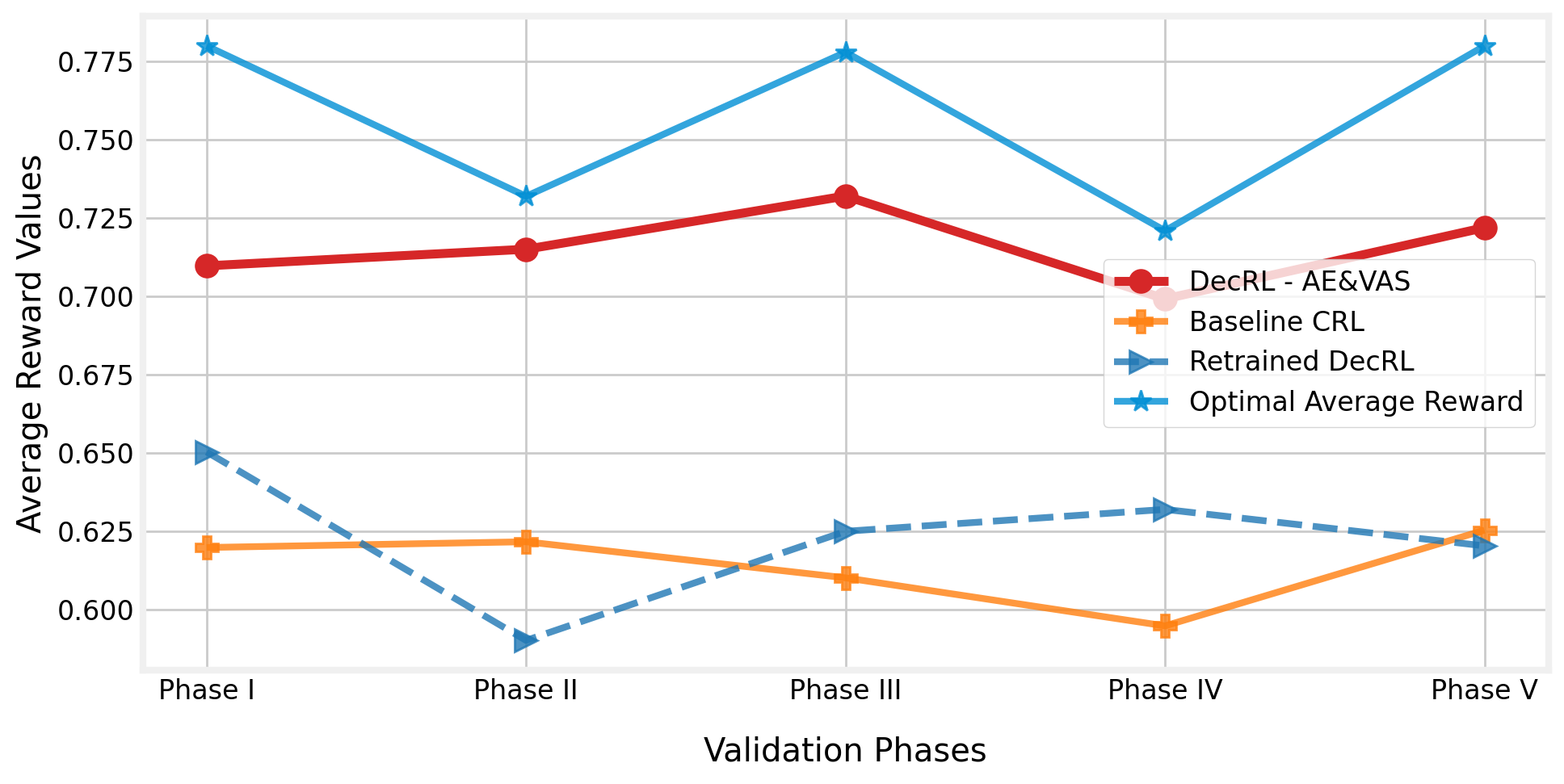}

\caption{Average Reward Value in each validation phase}

\label{fig:Avg_Reward}
\end{figure}

\begin{figure*}[t]
\centering
\includegraphics[scale=0.9]{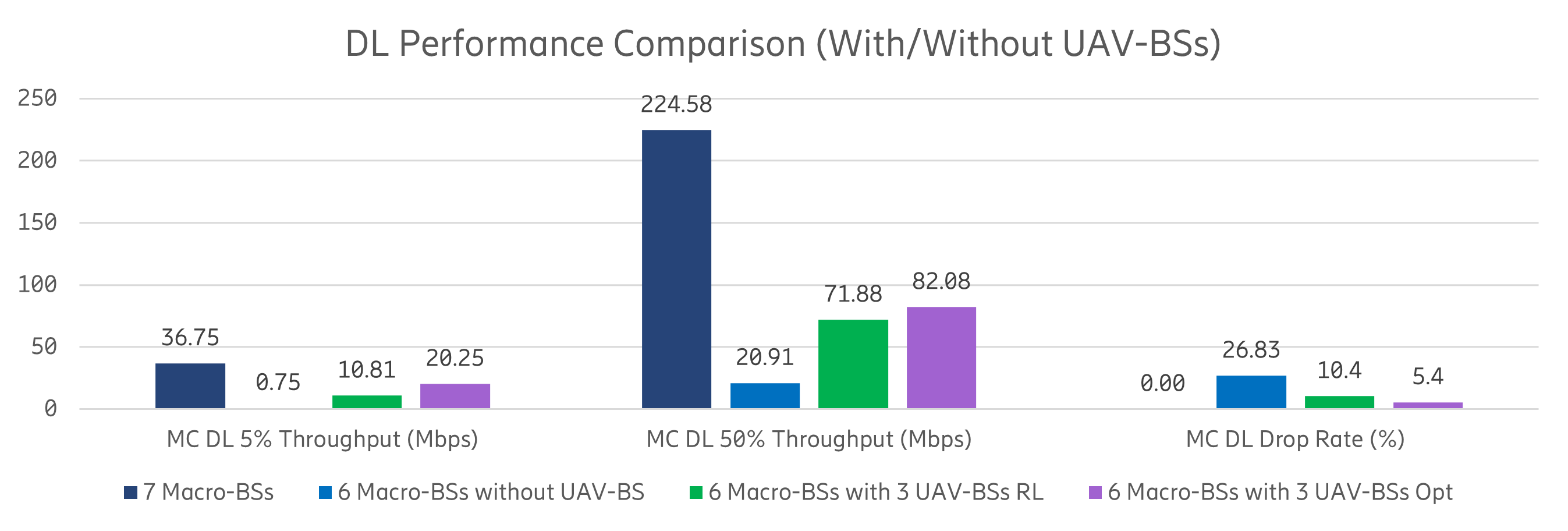}
\includegraphics[scale=0.9]{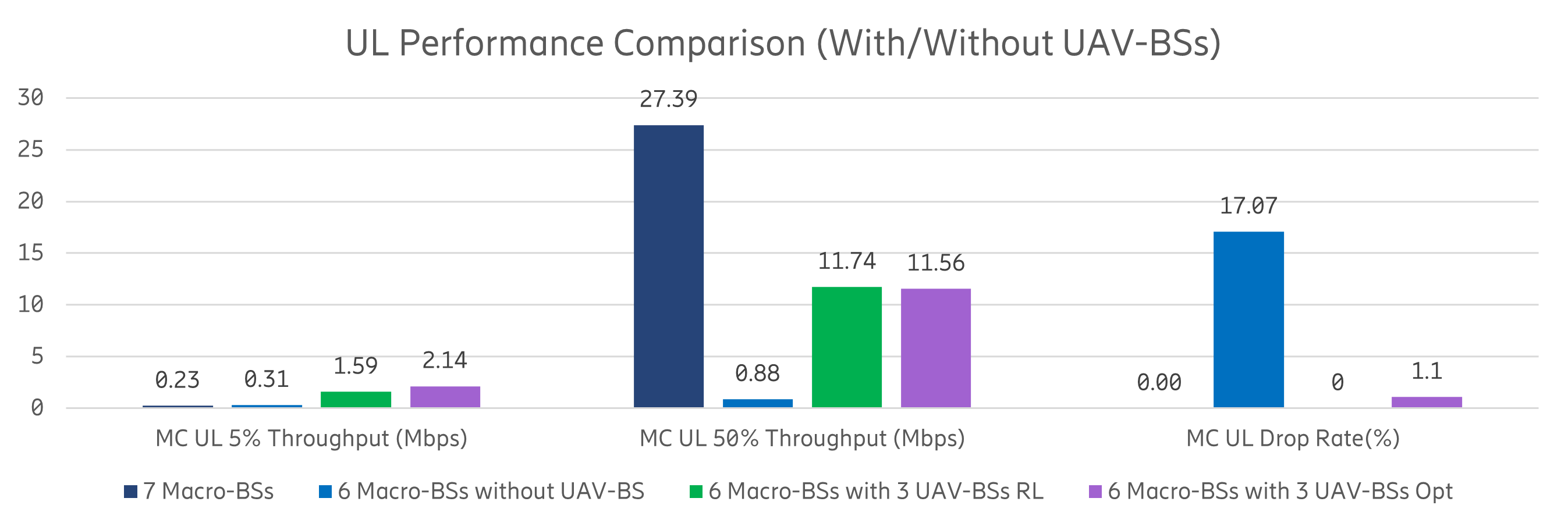}
\caption{DL \& UL performance comparison with/without UAV-BSs in the mission-critical use case}

\label{fig:per}
\end{figure*}

We first analyze the algorithm's convergence performance during the model training phase. Figure \ref{fig:rewardt} depicts the reward value as a function of the number of learning iterations. Compared to the baseline CRL algorithm, we can see that the VAS method can help the UAV-BS quickly discover the near-optimal position and converge at a high-quality level. 

Training the algorithm in a decentralized way can also be converged and reach a high reward value before 400 training rounds. The convergent reward value of the DecRL-AE\&VAS is approximately 5\% higher than that of the centralized training method in the initial learning iterations which results in higher training efficiency. When compared to the independent learning method, the algorithm does not converge after 1500 learning iterations. It is difficult for agents to share knowledge and collaborate since the algorithm stops them from exchanging information. Throughout the training procedure, the Baseline IRL algorithm performs poorly. Because the reward value represents the overall system performance of the three UAV-BSs, we can conclude that the DecRL-AE\&VAS algorithm enables the UAV-BSs to provide the best wireless connection service when compared to the other two frequently utilized baseline approaches in the training phase.

When entering the validation phases (Figure \ref{fig:Reward} Phase I - V), the proposed algorithm can learn from the past and finally reach the optimal state that provides the highest reward value in the validation scenarios using the AE method. Even if the environment has changed and the quality has dropped dramatically, the algorithm can assist the UAV-BS in quickly adjusting and returning to ideal performance. When we look at the baseline CRL approach, the method failed to respond to environmental changes in a short period due to the slow-paced state exploration. When we compared the baseline results to the retrained DecRL and DecRL-AE\&VAS, the results showed that our suggested VAS-AE approach can enable UAV-BS quickly converge in most of the validation phases. However, with the retrained DecRL, a significant impact may occur on the algorithm and service stability if not using previous existing knowledge. When we integrated RL with adaptive exploration and the value-based action selection strategy, the algorithm showed the best performance in terms of convergence speed, the ability to adapt to environmental changes and stable service quality of our suggested DecRL-AE\&VAS method.

The average reward value changes during the validation phases are depicted in Figure \ref{fig:Avg_Reward}. Because the reward value encompasses all of the essential criteria that must be evaluated during the deployment of the UAV-BS, the value clearly demonstrates the model's quality at each stage. As shown in the figure, our proposed DecRL-AE\&VAS method can assist the UAV-BS in maintaining the ideal service quality, but the baseline model failed to discover a state that can give reliable and satisfactory service to MC users. The suggested DecRL-AE\&VAS algorithm gives a reward value (a weighted sum of the six assessed performance measures) of only about 5\% to 6\% less than the global best solution in each validation phase.

In conclusion, we show that, when compared to the baseline RL approaches, the DecRL-AE\&VAS model can efficiently assist the UAV-BS in finding the near-optimal location and converging at a high-quality level. Utilizing the model-sharing and information-exchange technique, the proposed algorithm can learn from the past and eventually arrive at the optimal state that provides the best reward value, the proposed DecRL-AE\&VAS method can achieve the same or even higher levels of system quality than the Baseline CRL approach in both training and validation phases.  Because the reward value shows the system's performance, we may conclude that the DecRL-AE\&VAS can help UAV-BS provide better service than the other baseline models.  Our findings show that the algorithm has the capacity to link MC users swiftly and adequately, allowing drone systems to self-learn without centralized interaction. In Figure \ref{fig:per}, we also demonstrate the effectiveness after deploying multiple UAV-BSs in the MC use case. The "7 Macro-BSs" denotes the scenario where 7 marco-BSs are serving the whole scenario, while the "6 Macro-BSs without UAV-BS" denotes the scenario where one macro-BS is broken and no UAV-BS is deployed. The case "6 Macro-BSs with 3 UAV-BSs RL" describes the situation that 3 UAV-BSs are deployed to fill the coverage hole created in case "6 Macro-BSs without UAV-BS". The results in case "6 Macro-BSs with 3 UAV-BSs Opt" is derived based on grid search to be compared with the RL case. It can be observed that when one macro-BS brokes down, the MC users are experiencing severe performance degradation. Compared with the "6 Macro-BSs without UAV-BS" case, deploying 3 UAV-BSs as in the case "6 Macro-BSs with 3 UAV-BSs RL" can improve about 80\% system performance (including 5\%, 50\% throughput and drop rate) for MC users in both DL and UL. Furthermore, the suggested DecRL-AE\&VAS algorithm gives a throughput of 10 Mbps and a drop rate of only about 2\% to 3\% less than the global best solution indicated in the "6 Macro-BSs with 3 UAV-BSs Opt" case. In conclusion, deploying UAV-BS can greatly improve the performance of MC users who are experiencing coverage loss. However, due to the BS capability difference, deploying 3 UAV-BS, in this case, can not guarantee that the MC users have similar performance in the case when no emergency happens. The only exception is that the UL 5\% throughput performance of MC users served by 3 UAV-BSs is better than that served by one macro-BS before the disaster happens. This is because three UAV-BSs can be deployed in dispersed locations which increases the probability for an MC user to be close to its serving BS and the UL 5\% throughput performance of MC users is improved accordingly.

\section{Conclusions and Future Work}

In this paper, we presented a data collection system and machine learning applications for an MC use case, a novel RL algorithm, as well as a decentralized architecture to autonomously pilot multiple UAV-BS in order to offer users temporary wireless access. Two novel strategies, i.e., adaptive exploration and value-based action selection, are developed to help the proposed RL algorithms work efficiently in a dynamic real-world context, incorporating MC user movements and a decentralized architecture to support multi UAV-BSs deployment. Note that the number of participated UAV-BSs can be further extended based on the industrial requirements due to the characteristics of the decentralized architecture. We show that the proposed RL algorithm can monitor the MC service performance and quickly respond to environmental changes via self-adapting exploration probability. In addition, it requires far fewer model training iterations by reusing previous experiences and the value-based action selection strategy. Therefore, the proposed method can well serve the MC users by autonomously navigating multiple UAV-BSs despite environmental changes.

In the future, we will consider separating the configuration for access and backhaul antennas of the UAV-BS, as well as modelling drone rotation in the horizontal domain as an additional parameter for the UAV-BS configuration. We also intend to examine other hyper-parameters and reward function combinations based on different service requirements. Last but not least, we will also investigate the energy efficiency problem for deploying UAVs with machine learning components in the real-world context.


\bibliographystyle{IEEEtran}
\bibliography{cite}

\end{document}